\documentclass[superscriptaddress,runinaddress,showpacs,showkeys,aps,prd,floatfix,onecolumn]{revtex4}

\usepackage{eurosym}
\usepackage{amsfonts}
\usepackage{array}
\usepackage{amsthm}
\usepackage{bm}
\usepackage{palatino}
\usepackage[colorinlistoftodos]{todonotes}
\usepackage{mathpazo}
\usepackage{amssymb}
\usepackage{eurosym}
\usepackage{amsmath}
\usepackage{epsfig}
\usepackage{graphics}
\usepackage{color}
\usepackage{graphicx}
\usepackage[colorlinks=true,
            linkcolor=blue,
           urlcolor=blue,
           citecolor=blue]{hyperref}

\usepackage{changes}

\newcommand{\be}{\begin{equation}}
\newcommand{\ee}{\end{equation}}
\newcommand{\bea}{\setlength\arraycolsep{2pt} \begin{eqnarray}}
\newcommand{\eea}{\end{eqnarray}}

\begin{document}

\title{Shadow cast of non-commutative black holes in Rastall gravity}

\author{Ali \"{O}vg\"{u}n}
\email{ali.ovgun@emu.edu.tr}
\homepage[]{https://www.aovgun.com}
\affiliation{Instituto de F{\'i}sica, Pontificia Universidad Cat{\'o}lica de
Valpara{\'i}so, Casilla 4950, Valpara{\'i}so, Chile.}
\affiliation{Physics Department, Faculty of Arts and Sciences, Eastern Mediterranean
University, Famagusta, North Cyprus, via Mersin 10, Turkey}

\author{\.{I}zzet Sakall{\i}}
\email{izzet.sakalli@emu.edu.tr}
\affiliation{Physics Department, Faculty of Arts and Sciences, Eastern Mediterranean
University, Famagusta, North Cyprus, via Mersin 10, Turkey}

\author{Joel Saavedra}
\email{joel.saavedra@ucv.cl}
\affiliation{Instituto de F\'{\i}sica, Pontificia Universidad Cat\'olica de Valpara\'{\i}%
so, Casilla 4950, Valpara\'{\i}so, Chile}

\author{Carlos Leiva}
\email{cleivas62@gmail.com}

\affiliation{Departamento de F\'{\i}sica, Universidad de Tarapac\'{a}, Casilla 7-D, Arica, Chile}

\begin{abstract}
We study the shadow and energy emission rate of a spherically symmetric non-commutative black hole in Rastall gravity. Depending on the model parameters, the non-commutative black hole can reduce to the Schwarzschild black hole. Since the non-vanishing non-commutative parameter affects the formation of event horizon, the visibility of the resulting shadow depends on the non-commutative parameter in Rastall gravity. The obtained sectional shadows respect the unstable circular orbit condition, which is crucial for physical validity of the black hole image model. %Effect of the non-commutative parameter in Rastall gravity on the energy emission rate is also discussed.
\end{abstract}

\date{\today}

\maketitle

\section{Introduction}

The defining feature of a black hole (BH) is the event horizon, the boundary from within which a particle cannot escape. Often, this is described as the boundary within which the BH's escape velocity is greater than the speed of light. In other words, the gravitational pull becomes so large that none of the physical particles can escape, including light. On the other hand, outside the event horizon, light can escape from a BH. The surrounding matter pulled by BH is called accretion. The accretion overheats in time because of viscous dissipation. Thus, it radiates very brightly at different frequencies including the radio waves, which can be detected by radio telescopes. This shining material accreting onto the BH crosses the event horizon, resulting in a dark area over a bright background: this is the so-called BH shadow (BHS). Although the concept of BHS was known since seventies, the idea to image the BHS locating in the center of our Milky Way was first considered by Falcke et al (2000) \cite{Falcke:1999pj}. Recently, Event Horizon Telescope has observed the image of BHS of Messier 87 galaxy \cite{Akiyama:2019eap}. With this observation, the shadow of BHS has become a very popular subject in today's literature \cite{synge,luminet,cunha,Ovgun:2018tua,Kuniyal:2017tue,Wei:2015dua,Saha:2018zas,Maceda:2018zim,Cunha:2018acu,Tremblay:2016ijg,Shen:2005cw,Huang:2007us,Johannsen:2015mdd,Konoplya:2019sns,Shaikh:2019fpu,Wang:2018prk,Cunha:2018gql,Hioki:2009na,Johannsen:2010ru,Nedkova:2013msa,Amarilla:2013sj,Abdujabbarov:2012bn,Abdikamalov:2019ztb,Abdujabbarov:2017pfw,Abdujabbarov:2016efm,Abdujabbarov:2016hnw,Papnoi:2014aaa,Atamurotov:2013sca,Grenzebach:2014fha,Johannsen:2015hib,Giddings:2014ova,Wei:2013kza,Sakai:2014pga,Perlick:2015vta,Jusufi:2017mav,Abdujabbarov:2015xqa,Tinchev:2013nba,Wang:2018eui,Amarilla:2010zq,Yumoto:2012kz,Takahashi:2005hy,Dexter:2012fh,Younsi:2016azx,Johannsen:2015qca,Freivogel:2014lja,Cunha:2016bpi,Ohgami:2015nra,Zakharov:2011zz,Hennigar:2018hza,Pu:2016qak,Sharif:2016znp,Xu:2018mkl,Gyulchev:2018fmd,Vetsov:2018mld,Perlick:2018iye,Stuchlik:2018qyz,Shaikh:2018lcc,Eiroa:2017uuq,Mars:2017jkk,Wang:2017hjl,Tsukamoto:2017fxq,Singh:2017vfr,Mureika:2016efo,Huang:2016qnl,Tsukamoto:2014tja,Kumar:2017vuh,Kumar:2018ple,Giddings:2019jwy,Giddings:2019vvj,Contreras:2019cmf,Contreras:2019nih}.
 
Despite intense efforts over the last years, it is still far from being understood what a consistent theory of quantum gravity will look like and what its main features will be. There are various number of proposals to find an effective quantum gravity theory such as loop quantum gravity, string theory, non-commutative geometry etc.  On the other hand, BHs have curvature singularities and to heal this problem, quantum gravity can be used where general relativity not anymore working well at small scales \cite{Jusufi:2015mii,Kuang:2017sqa}. Non-commutative (NC) theory can modify the structure of spacetime by assuming universal minimal length scale where  quantization of spacetime \cite{Hod:2018xfy,Sakalli:2014wja,Sakalli:2016fif,Sakalli:2018nug} needs non commute of its coordinates :
\begin{equation*}
[x^{A},x^{B}]=i \vartheta^{AB},
\end{equation*}
in which $\vartheta^{AB}$ is a real-valued anti-symmetric matrix with
$\vartheta^{AB}=\vartheta \text{diag}(\epsilon_{ij},~\epsilon_{ij},...)$,
$\vartheta$ is a constant of dimension $\text{[Length]}^{2}$. In the
limit $\vartheta\rightarrow 0$, the ordinary spacetime is recovered.
Physically, $\vartheta^{AB}$ represents a small patch in $AB$-plane of
the observable area as the Planck's constant ($\hbar$) illustrates
the smallest fundamental cell of the observable phase space in
quantum mechanics.

Using the NC spacetimes, one can remove the singularities from the BHs and construct regular BH. Afterwards, many other important aspects of
spacetime non-commutativity have been discussed in literature [see for example \cite{Nicolini:2005vd,Nicolini:2008aj,Ansoldi:2006vg,Banerjee:2008gc,Spallucci:2009zz,Rizzo:2006zb,Nouicer:2007jg,Modesto:2010rv,Nicolini:2009gw,Balart:2014cga,Ma:2017jko,Sadeghi:2018vrf}].
 
In this paper, we study the shadow of charged non-rotating NC geometry inspired Schwarzschild BH (NCGiSBH) belonging to the Rastall gravity. To this end, we consider both equatorial and non-equatorial planes and depict the shadow with the help of celestial coordinates. Furthermore, we compute the energy emission rate of the NC BH and discuss the energy influence from the plasma of accretion on the shadow. 

The paper is organized
as follows. In the next section, we introduce the NC BH in the Rastall gravity. Section \ref{sec3} is  devoted to study the apparent shape
of the NCGiSBH geometry by finding its shadow.
We also investigate the influence of plasma on the shadows of the NCGiSBH. Our final remarks are presented in Sect. \ref{sec4}.

\section{NC BH spacetime in Rastall gravity} \label{sec2}

NC BH arises as solutions of Einstein equation where the coordinates are NC and satisfy the following algebra $[x^{\mu},x^{\nu}]=i\vartheta^{\mu\nu}$ in which $\vartheta^{\mu\nu}$ represents an anti-symmetric tensor encoded the discretization of the spacetime.

Using the NC algebra, one can construct a spacetime, having Schwarzschild BH limit \cite{Nicolini:2005vd} as follows: 
\begin{equation}
ds^2 = - f(r)  \hspace{0.1cm} dt^2 + \frac{dr^2}{f(r)} + r^2  \hspace{0.1cm}[ d\theta^2 + sin^2(\theta) \hspace{0.1cm}d\phi^2 ],
\label{BH1}
\end{equation}
where,
\begin{equation}
f(r) = 1 - \frac{4 \hspace{0.1cm} M}{r  \hspace{0.1cm}\sqrt{\pi}} \gamma\left(\frac{3}{2}, \frac{r^2}{4  \hspace{0.1cm}\epsilon}\right),
\label{gamma}
\end{equation}
with the gamma $\gamma$  function:
\begin{equation}
\gamma\left(\frac{3}{2}, \frac{r^2}{4 \epsilon}\right) \equiv \int_{0}^{\frac{r^2}{4 \epsilon}} \sqrt{t} \hspace{0.1cm} e^{-t}  \hspace{0.1cm}dt.
\end{equation}
Here, $\sqrt{\epsilon}$ stands for the width of the Gaussian mass-energy. The NC BH reduces to the Schwarzschild BH as  $\frac{r}{\sqrt{\epsilon}} \rightarrow \infty$. The horizon of the BH is conditional on $f(r_h) = 0$:
\begin{equation}
 r_h = \frac{4 \hspace{0.1cm} M}{ \sqrt{\pi}} \gamma\left(\frac{3}{2}, \frac{r^2_{h}}{4  \hspace{0.1cm}\epsilon}\right).
\label{horizon}
\end{equation}

Note that the NC effects are dominant at short distances, on the other hand, at large distances, they become non-effective.

In Rastall gravity theory, the source of matter diffuses on the region. In this theory, the mass density of the smeared matter distribution and the corresponding ansatz of the NC BH metric are given by \cite{Ma:2017jko}
\be
\rho =-T^t_{~t}=\frac{M }{(4 \pi  \vartheta)^{3/2}}\exp \left(-\frac{r^2}{4 \vartheta}\right),
\ee

\be\label{fr1}
f(r)=B_1-\frac{B_0}{r}-\frac{2GM}{r\sqrt{\pi}} \gamma\left(\frac{1}{2},\frac{r^2}{4\vartheta}\right). 
\ee
It is noted that the gamma function seen in Eq. \eqref{fr1} is given by
  \be\label{sol00}
     \gamma\left(\frac{1}{2},\frac{r^2}{4\vartheta}\right)=\int_0^{r^2/4\vartheta}dt~ t^{-1/2}e^{-t}.
    \ee
The integration constants are fixed at $B_1=1$ and $B_0=0$ so that metric function becomes

\be\label{sol1}
f(r)\equiv f= 1 - \frac { 2 G M } { r \sqrt { \pi } } \gamma\left(\frac{1}{2},\frac{r^2}{4\vartheta}\right),
\ee
where the solution \eqref{sol1} asymptotically approaches to the Schwarzschild BH since ($\gamma(1/2,r^2/4\vartheta) \rightarrow \sqrt{\pi}$). For this reason, we call the metric \eqref{BH1} with Eq. \eqref{sol1} as the NCGiSBH (in the
Rastall gravity). The NCGiSBH solution is finite at the origin:
\be
f=1-\frac{2GM}{\sqrt{\pi\vartheta}}+\frac{GM r^2}{6 \sqrt{\pi } \vartheta ^{3/2}}+O(r^4),
\ee
but it is a singular BH, which can be best seen from its Ricci scalar:
\be 
R=\frac{G M e^{-\frac{r^{2}}{4 \vartheta}}\left(4 \vartheta-r^{2}\right)}{\sqrt{\pi} \vartheta^{3 / 2} r^{2}}.
 \ee

When $M/\sqrt{\vartheta}<\sqrt{\pi}/2$, the NCGiSBH does not possess an event horizon. On the other hand,  it has an event horizon $r{_H}$ when $M/\sqrt{\vartheta}>\sqrt{\pi}/2$. The Hawking temperature \cite{newwald} of the NCGiSBH can be found as
\be
T_{H}=\frac{f^{\prime}\left(r_{H}\right)}{4 \pi}=\frac{1}{4 \pi r_{H}}\left[1-\frac{r_{H} e^{-\frac{r_{H}^{2}}{4 \vartheta}}}{\sqrt{\vartheta} \gamma\left(1 / 2, \frac{r_{H}^{2}}{4 \vartheta}\right)}\right]. \label{TH}
 \ee 
 It is worth noting that as it was stated in \cite{Ma:2017jko}, while $r_{H}/\sqrt{\vartheta}$ gets large values, the temperature of the NCGiSBH almost matches with the temperature of the Schwarzschild BH. It was thoroughly discussed that although the NCGiSBH of GR theory admits a zero-temperature remnant with
radius $r_{H} = 3\sqrt{\vartheta}$, the NCGiSBH of the Rastall gravity evaporates by ending up a point-like ($r_{H} = 0$) massive remnant. From now on, in our computations we shall focus on the NCGiSBH of the Rastall gravity.

\section{Null Geodesics and Shadow Casts of NCGiSBH in Rastall gravity} \label{sec3}

For studying the BHS, it is necessary to consider the geodesics of a test particle in the associated BH geometry. To this end, we use the Hamilton-Jacobi and Carter's constant separable methods for the NCGiSBH geometry. 

The relativistic Hamilton-Jacobi equation reads \cite{Singh:2017xle}

\begin{equation}
\frac{\partial S}{\partial \sigma}=-\frac{1}{2}g^{\mu\nu}\frac{\partial S}{\partial {x^{\mu}}}\frac{\partial S}{\partial {x^{\nu}}},
\label{HJ1}
\end{equation}
where $S$ is called the Jacobi action and $\sigma$ denotes the affine parameter along the geodesics. The separable solution for the Jacobi action $S$ can be introduced as \cite{Abdujabbarov:2016hnw,Hou:2018bar}
\begin{equation}
S=\frac{1}{2}m^2\sigma-Et+L\phi+S_r(r)+S_{\theta}(\theta),
\label{HJ2}
\end{equation}
where $m$, $E$, and $L$ are the test particle's mass, energy, and angular momentum, respectively. $S_r(r)$ and $S_{\theta}(\theta)$ are functions of $r$ and $\theta$, respectively. Inserting Eq. (\ref{HJ2}) into Eq. (\ref{HJ1}) and applying the method of separation of variables, one can obtain the null geodesic (for photon $m=0$) equations for a test particle around the NCGiSBH as follows
\begin{align}
\label{HJ3}
&r^2\frac{dt}{d\sigma}=\frac{1}{f}(Er^2),\\
&r^2\frac{dr}{d\sigma}=\sqrt{\mathcal{R}},\\
&r^2\frac{d\theta}{d\sigma}=\sqrt{\Theta},\\
\label{HJ4}
&r^2\frac{d\phi}{d\sigma}=\left(\frac{L}{\sin^2\theta}\right),
\end{align}
where $\mathcal{R}(r)$ and $\Theta(\theta)$ take the following forms
\begin{align}
\label{HJ5}
&\mathcal{R}(r)=E^2r^4-r^2 f[m^2r^2+L^2+\mathcal{K}],\\
&\Theta(\theta)=\mathcal{K}-\left(  \dfrac{L^2}{\sin^2\theta}  \right) \cos^2\theta,
\end{align}
in which $\mathcal{K}$ is the Carter constant \cite{Johannsen:2015mdd}. The above geodesic equations (\ref{HJ3}-\ref{HJ4}) fully describe the dynamics of the test particle around the NCGiSBH. The boundary of the BHS is mainly determined by the unstable circular orbit. Here,  we consider the case of photons ($m=0$). For an observer afar off a BH, the photons arrive the BH near the equatorial plane ($\theta=\pi/2$).  The unstable circular orbits satisfy the following condition:
\begin{equation}
\mathcal{R}(r_c)=\frac{\partial\mathcal{R}(r_c)}{\partial r}=0.
\label{R1}
\end{equation}
Note that $r_c$ is an important parameter for determing shape of BHS where for Schwarzschild black hole, $r_c = 3M$ for the equatorial circular orbit. Moreover, it is impossible to obtain $r_c$ for the non-commutative black hole \cite{Wei:2015dua}.

Recalling Eq. (\ref{HJ5}) and by introducing the two impact parameters $\xi$ and $\eta$ :
\begin{equation}
\xi=L/E,  \;\; \;\; \;\;   \eta=\mathcal{K}/E^2,
\end{equation}
we obtain

\begin{equation}
\label{R2}
r_c^4-[\eta+\xi^2][r_c^2 f(r_c)]=0,
\end{equation}
\begin{equation}
\label{R3}
4r^3-[\eta+\xi^2][2r_c f(r_c) +r_c^{2}{f(r_c)}2 f'(r_c)]=0. \end{equation}

Combining Eqs. (\ref{R2}-\ref{R3}) with Eq. \eqref{sol1} and making some algebra, $\xi^2+\eta$ yields

\begin{equation}
\label{R44}
\eta+\xi^{2}=\frac{ r_c^{2}}{f(r_c)}
\end{equation}
 and get the relation for the photon sphere radius $r_c$ as follows: \begin{equation}
\frac{f^{\prime}(r_c)}{f(r_c)}=\frac{2}{r_c}.
\end{equation}
One can find the shadow size
from \ref{R44}, after putting this photon sphere radius $r_c$.

\begin{figure}
\begin{center}
    \includegraphics[scale=0.3]{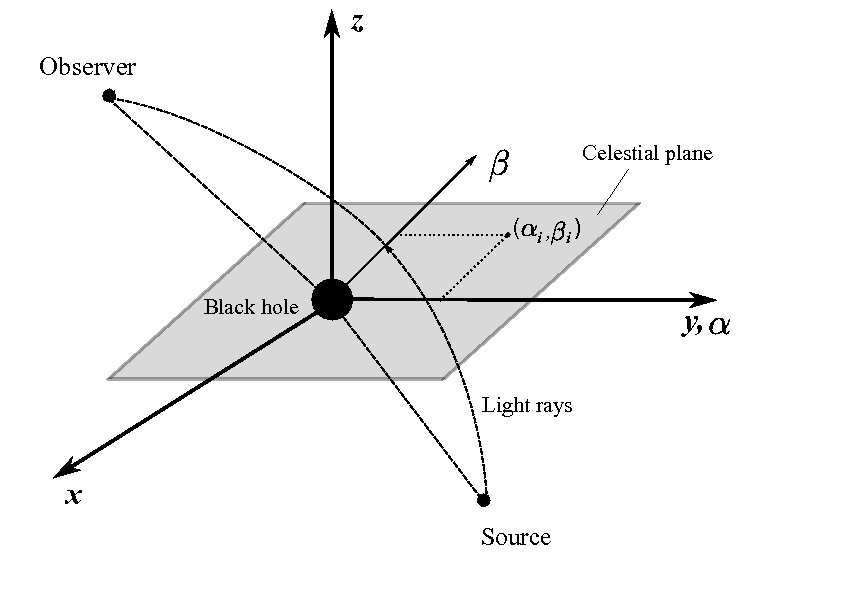}
    \end{center}
\caption{Schematic illustration \cite{Abdujabbarov:2016hnw}}
       \label{fig:0}
\end{figure}

To determine the shape of the BHS, we introduce the following celestial coordinates ($\alpha$ and $\beta$) :
\begin{align}
& \alpha = \lim_{r_c\to \infty}\left( -r_c^2 \sin \theta_0 \dfrac{d\phi}{dr}  \right),\\
& \beta = \lim_{r_c \to \infty}\left( r_c^2 \dfrac{d\theta}{dr}  \right),
\end{align}
where $r_o$ is the distance between the BH and the observer and $\theta_o$, inclination angle, represents the angle between the rotation axis of the BH and the line of sight of the observer as shown in Fig. \ref{fig:0}. $\alpha$ measures the apparent perpendicular distance of the shadow as seen from the axis of symmetry and $\beta$ reads the apparent perpendicular distance of the shadow as seen from its projection on the equatorial plane.

Using the null geodesic Eqs. (\ref{HJ3}-\ref{HJ4}), we can obtain the relations between celestial coordinates and impact parameters, $\xi$ and $\eta$, as follows
\begin{align}
& \alpha = -\dfrac{\xi}{sin \theta},\\
& \beta = \pm \sqrt{\eta  -\xi^2\cot^2 \theta}.
\end{align}

In the equatorial plane ($\theta=\pi/2$), $\alpha$ and $\beta$ reduce to
\begin{align}
& \alpha = -\xi,\\
& \beta = \pm \sqrt{\eta }.
\end{align}

 %The shadow of the NCGiSBH is nothing but a perfect circle with radius of $R_s$ \cite{Singh:2017xle}.  %Thus, we have
%\begin{equation}
%\alpha^2 + \beta^2 = \xi^2 +\eta= R_s^2. 
%\end{equation}

\subsection{Shadow cast and energy emission rate of NCGiSBH: Observable quantities}

\begin{figure}
    \begin{center}
    \includegraphics[scale=0.20]{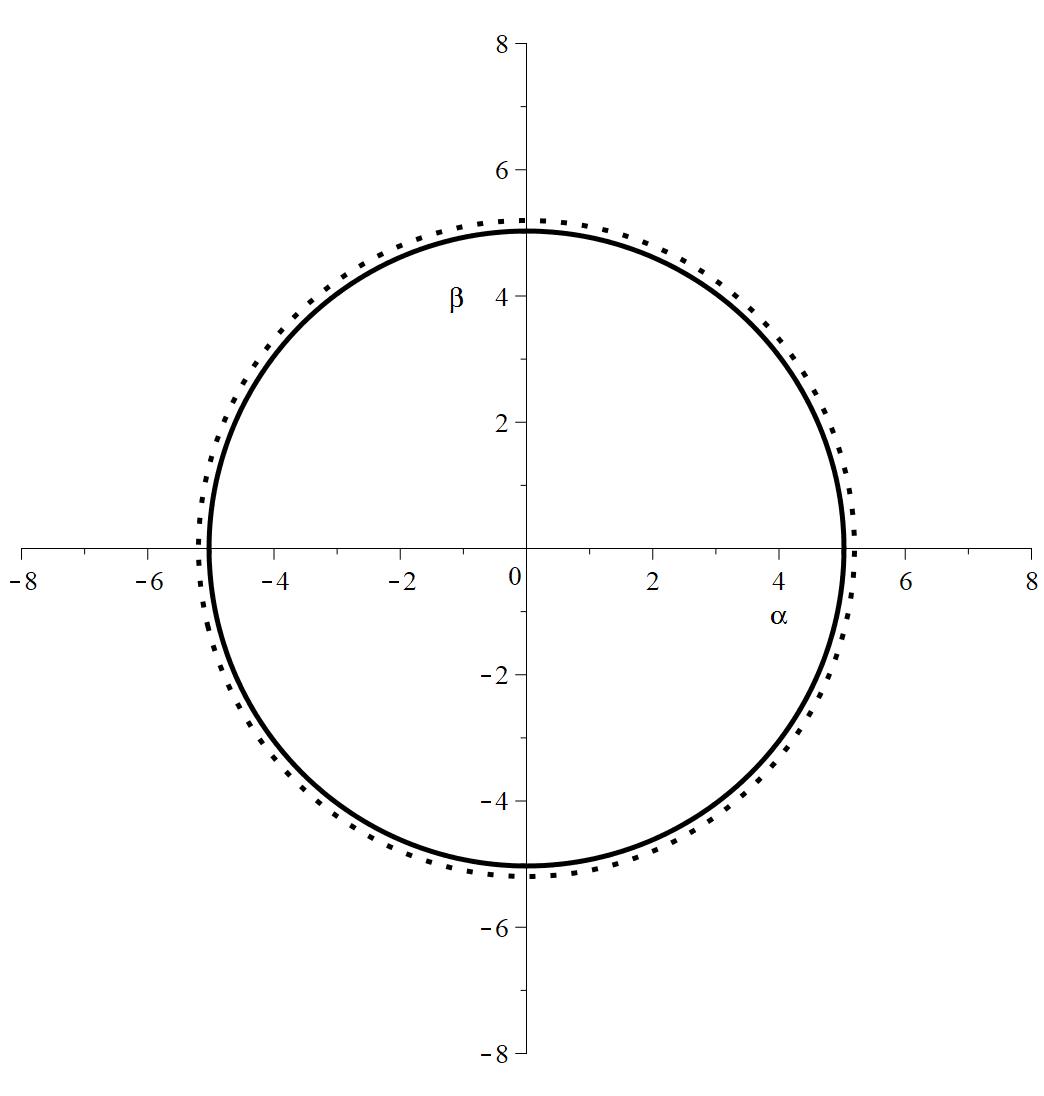}
    \end{center}
\caption{ $M=1$ and the dot line is for $\vartheta=0.001$ and the solid line is for $\vartheta=0.999$. }
       \label{fig:1}
\end{figure}

Since photons coming from both sides of the static BH have the same value of the
deflection angle, the shadow of such case is nothing but a standard circle with radius $R_{s}/M=\sqrt{\alpha^{2}+\beta^{2}}=\sqrt{\xi^2 +\eta}$ \cite{Wei:2015dua}. 
Using the radii of the BHS are presented in Fig. (\ref{fig:1}). 

\begin{figure}
\begin{center}
    \includegraphics[scale=0.60]{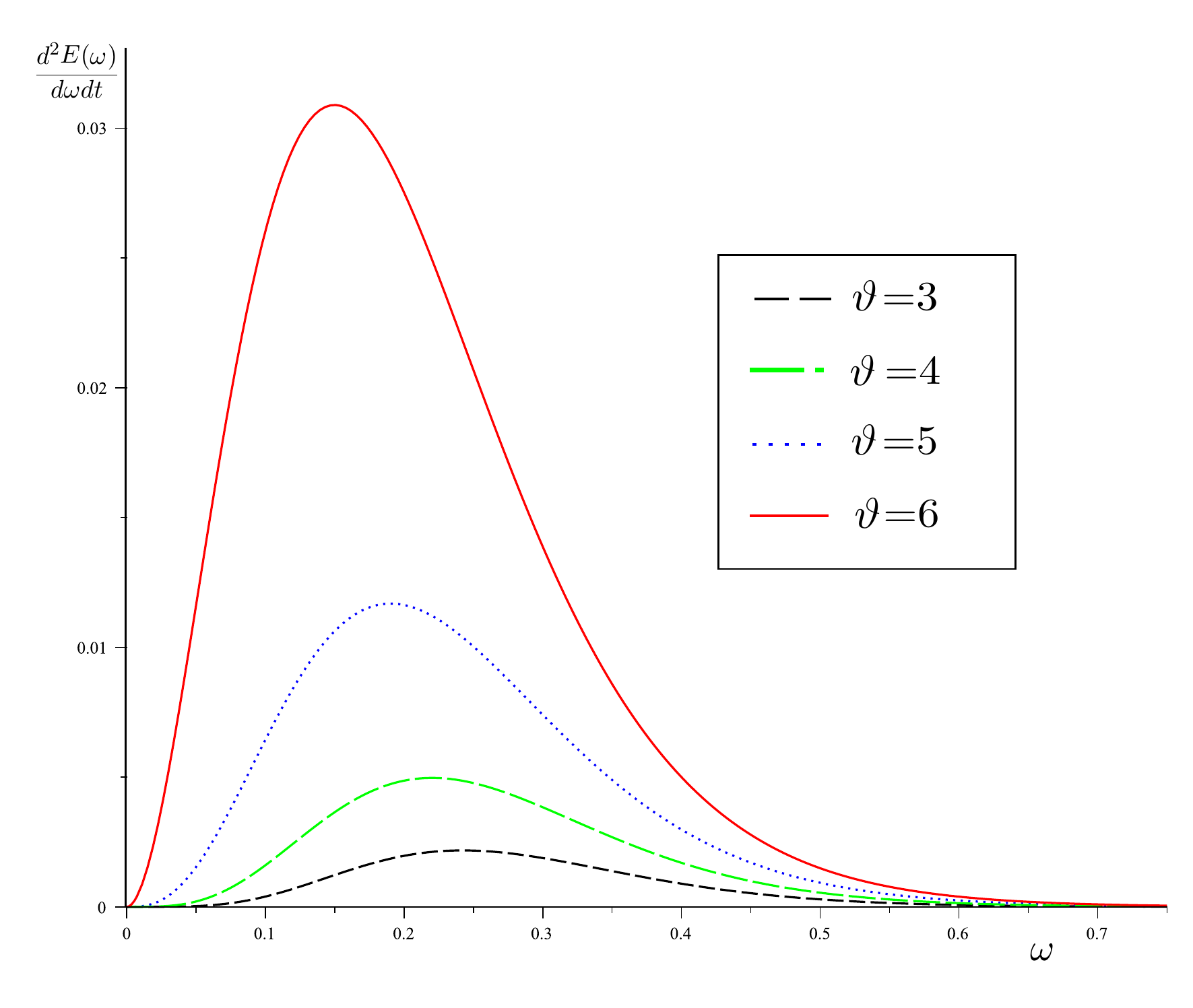}
    \end{center}
\caption{Plot showing changing of energy emission rate with frequency $\vartheta$. The other physical parameters are chosen as $r_{0}=0.3$ and $r_{h}=1$.}
       \label{fig:3}
\end{figure}

For an observer located at an infinite distance, the BHS corresponds to the high energy absorption cross section, which oscillates around geometrical optics limit value $\sigma_{lim}$ for a spherical symmetric BH \cite{testlim}. $\sigma_{lim}$ is approximately equal to the geometrical cross section of the photon sphere:  
\begin{equation}
\sigma_{lim} \approx \pi R_s^2,
\end{equation}
where $R_s$ is the radius of the BHS. On the other hand,the energy emission rate of the black hole is given by \cite{emis} 
\begin{equation}
\dfrac{d^2E(\omega)}{d\omega dt} = \dfrac{2\pi^2\sigma_{lim} }{e^{\omega/T_{H}}-1}\omega^3,
\end{equation}
where $\omega$ is the frequency of photon and recall that $T_{H}$ is the Hawking temperature \eqref{TH} of the NCGiSBH. In Fig. (\ref{fig:3}), we depict the energy emission rate with frequency $\omega$ for the NCGiSBH. It can be seen that the peak value of that Gaussian type plot decreases when $\vartheta$ gets higher values. The latter remark ad-signifies that the NCGiSBH's area will also decrease for the higher values of $\vartheta$.

\section{Conclusions} \label{sec4}

In this study, we have studied the shadow of the NCGiSBH in the Rastall gravity. We have shown that for the NCGiSBH, the shape of the BHS is nothing but a ring and its radius depends on $\sqrt{\vartheta}M^{-1}$.   
    The Fig. (\ref{fig:1}) implies that the shadow radius decreases with $\sqrt{\vartheta}M^{-1}$ increases.
    Therefore, the effect of the NC parameter in Rastall gavity varies the size of the shadow. 
    We have also discussed the energy emission rate, which is another observable phenomenon, for the NCGiSBH. As can be seen from Fig. (\ref{fig:3}), the obtained Gaussian type graph decreases its peak value as $\sqrt{\vartheta}M^{-1}$ increases. For this reason, the area of NCGiSBH shrinks with increasing $\sqrt{\vartheta}M^{-1}$ value.
     
    The observational study of the BHS has just initiated with Event Horizon Telescope (EHT) \cite{EHT}. In fact, the image of EHT does not depict the BH's event horizon, but a shadow cast by the light around it due to the unstable orbits of photons around the central object. We believe that in the near future many other images belonging to the various BHSs will be available due to the developing technology and science. Thus, we are at the forefront of the BHS and this subject will be a test center for GR in the forthcoming years. In this context, the determination of what kind of BH is observed will be of great importance. Similar to this study, in the near future, by using the geometric optics, we will try to identify the differences between the BHs obtained from different theories.

  \acknowledgments
We wish to thank the Editor and anonymous
Reviewers for their valuable comments and suggestions. The work of C. L. was supported in part by the UTA Mayor 4736-18. This work is supported by Comisi{\'o}n Nacional de
Ciencias y Tecnolog{\'i}a of Chile (CONICYT) through FONDECYT Grant N{$\mathrm{o}$} 3170035 (A. {\"O}.).

\end{document}